\documentclass[aps, a4paper, twocolumn, superscriptaddress, amsmath]{revtex4-1}

\usepackage{epsf, latexsym, color, epsfig, ulem}
\usepackage{amssymb, amsmath}
\usepackage{times}
\usepackage{qcircuit}
\usepackage{verbatim}
\usepackage{booktabs}
\usepackage{multirow}
\usepackage{gensymb}
\usepackage{xcolor}

\usepackage[colorlinks]{hyperref}
\AtBeginDocument{%
    \hypersetup{
      urlcolor     = blue, 
      filecolor   = red,
      linkcolor    = blue, 
      citecolor   = blue, 
    }
} 

\newcommand{\ket}[1]{| #1 \rangle}

\newcommand{\etal}{{\textit{et al.}}}

\newcommand\diag{{\mbox{diag\,}}}

\newcommand{\ignore}[1]{}

\definecolor{darkgreen}{rgb}{0,0.5,0}
\definecolor{darkred}{rgb}{0.5,0,0}
\definecolor{darkblue}{rgb}{0.0,0,0.5}

\newcommand{\be}{\begin{equation}}
\newcommand{\ee}{\end{equation}}
\newcommand{\ba}{\begin{eqnarray}}
\newcommand{\ea}{\end{eqnarray}}
\newcommand{\bc}{\begin{center}}
\newcommand{\ec}{\end{center}}

\def\CC{{\rm\kern.24em \vrule width.04em height1.46ex depth-.07ex
    \kern-.30em C}}
\def\P{{\rm I\kern-.25em P}}

\def\RR{{\rm
         \vrule width.04em height1.58ex depth-.0ex
         \kern-.04em R}}

\def\bbbc{{\mathchoice {\setbox0=\hbox{$\displaystyle\rm C$}\hbox{\hbox
to0pt{\kern0.4\wd0\vrule height0.9\ht0\hss}\box0}}
{\setbox0=\hbox{$\textstyle\rm C$}\hbox{\hbox
to0pt{\kern0.4\wd0\vrule height0.9\ht0\hss}\box0}}
{\setbox0=\hbox{$\scriptstyle\rm C$}\hbox{\hbox
to0pt{\kern0.4\wd0\vrule height0.9\ht0\hss}\box0}}
{\setbox0=\hbox{$\scriptscriptstyle\rm C$}\hbox{\hbox
to0pt{\kern0.4\wd0\vrule height0.9\ht0\hss}\box0}}}}

\def\bbbq{{\mathchoice {\setbox0=\hbox{$\displaystyle\rm Q$}\hbox{\raise
0.15\ht0\hbox to0pt{\kern0.4\wd0\vrule height0.8\ht0\hss}\box0}}
{\setbox0=\hbox{$\textstyle\rm Q$}\hbox{\raise
0.15\ht0\hbox to0pt{\kern0.4\wd0\vrule height0.8\ht0\hss}\box0}}
{\setbox0=\hbox{$\scriptstyle\rm Q$}\hbox{\raise
0.15\ht0\hbox to0pt{\kern0.4\wd0\vrule height0.7\ht0\hss}\box0}}
{\setbox0=\hbox{$\scriptscriptstyle\rm Q$}\hbox{\raise
0.15\ht0\hbox to0pt{\kern0.4\wd0\vrule height0.7\ht0\hss}\box0}}}}

\def\bbbt{{\mathchoice {\setbox0=\hbox{$\displaystyle\rm
T$}\hbox{\hbox to0pt{\kern0.3\wd0\vrule height0.9\ht0\hss}\box0}}
{\setbox0=\hbox{$\textstyle\rm T$}\hbox{\hbox
to0pt{\kern0.3\wd0\vrule height0.9\ht0\hss}\box0}}
{\setbox6=\hbox{$\scriptstyle\rm T$}\hbox{\hbox
to0pt{\kern8.3\wd0\vrule height0.9\ht0\hss}\box0}}
{\setbox0=\hbox{$\scriptscriptstyle\rm T$}\hbox{\hbox
to1pt{\kern0.3\wd1\vrule height0.9\ht0\hss}\box0}}}}

\def\bbbz{{\mathchoice {\hbox{$\sf\textstyle Z\kern-0.4em Z$}}
{\hbox{$\sf\textstyle Z\kern-0.4em Z$}}
{\hbox{$\sf\scriptstyle Z\kern-0.3em Z$}}
{\hbox{$\sf\scriptscriptstyle Z\kern-0.2em Z$}}}}

\newcommand{\putfig}[2]{$$\leavevmode\hbox{\epsfxsize=#2 cm
   \epsffile{#1}}$$}

\begin{document}

\title{Interaction-free imaging of multi-pixel objects}

\author{Alexandra Maria P\u alici}
\email[Corresponding author: ]{a.palici@theory.nipne.ro}
\affiliation{Horia Hulubei National Institute of Physics and Nuclear Engineering, 077125 Bucharest--M\u agurele, Romania}

\author{Tudor-Alexandru Isdrail\u a}
\affiliation{Horia Hulubei National Institute of Physics and Nuclear Engineering, 077125 Bucharest--M\u agurele, Romania}

\author{Stefan Ataman}
\affiliation{Extreme Light Infrastructure-Nuclear Physics (ELI-NP), Horia Hulubei National Institute for Physics and Nuclear Engineering, 30 Reactorului Street, 077125 Bucharest--M\u agurele, Romania}

\author{Radu Ionicioiu}
\email[Corresponding author: ]{r.ionicioiu@theory.nipne.ro}
\affiliation{Horia Hulubei National Institute of Physics and Nuclear Engineering, 077125 Bucharest--M\u agurele, Romania}

\begin{abstract}
Quantum imaging, one of the pillars of quantum technologies, is well-suited to study sensitive samples which require low-light conditions, like biological tissues. In this context, interaction-free measurements (IFM) allow us infer the presence of an opaque object without the photon interacting with the sample. Current IFM schemes are designed for single-pixel objects, while real-life samples are structured, multi-pixel objects. Here we extend the IFM imaging schemes to multi-pixel, semi-transparent objects, by encoding the information about the pixels into an internal degree of freedom, namely orbital angular momentum (OAM). This allows us to image the pixels in parallel. Our solution exhibits a better theoretical efficiency than the single-pixel case. Our scheme can be extended to other degrees of freedom, like the photon radial quantum number, in order to image 1D and 2D objects. 

\end{abstract}

\maketitle

\section{Introduction}

Quantum imaging \cite{DAngelo2005, Genovese2016, Moreau2019} established itself as a major pillar of the rapidly expanding field of quantum technologies \cite{Dowling2015, Acin2018}. Among the techniques found in quantum imaging we can cite ghost imaging \cite{Pittman1995, Malik2010, Shapiro2012, Padgett2017}, imaging with undetected photons \cite{Lemos2014, Kalashnikov2016, Paterova2020, Kviatkovsky2020, PaterovaScience2020}, sub-shot-noise quantum imaging \cite{Treps2002, Brida2010, Samantaray2017} and finally, interaction-free measurement (IFM) imaging \cite{white98, Peise2015}. These quantum imaging techniques are not necessarily separated, as shown in reference \cite{Zhang2019, hance2021}, where interaction-free and ghost imaging techniques have been jointly used to harvest their full potential.

Ghost, or coincidence imaging, uses correlated light beams to image objects \cite{Pittman1995, Malik2010, Shapiro2012, Padgett2017}. The beam intersecting the object is collected into a ``bucket detector'' giving no spatial resolution. The image is then recovered by coincidences between the bucket detector and the free propagating beam that never intersected the object, hence the name ``ghost'' imaging \cite{Shapiro2008,Shapiro2012}. One can distinguish between quantum and classical correlation ghost imaging \cite{Gatti2004}, the advantage of the quantum ghost imaging being the ability to image objects both in the near and far field \cite{Bennink2004}.

The technique of imaging with undetected photons stemmed from the 1991 ``induced coherence'' experiment \cite{Wang1991}. More than two decades after this ``mind-boggling experiment'' \cite{Greenberger1993}, the now famous imaging with undetected photons was performed \cite{Lemos2014}. Further developments included infrared spectroscopy with visible photons \cite{Kalashnikov2016},  applications to semiconductor industry \cite{Paterova2020} and biological microscopy with undetected photons \cite{Buzas2020, Kviatkovsky2020, PaterovaScience2020}.

IFM imaging is based on a counter-intuitive feature of quantum mechanics, first noticed by Reinginer \cite{Renninger1960}, namely the concept of ``negative results'' of a measurement. The concept was further refined by Dicke \cite{Dicke1981} and finally, Elitzur and Vaidman \cite{EV} presented their celebrated bomb tester in 1993, based on a Mach-Zehnder (MZI) interferometer. Although it created a stir \cite{duMarchievanVoorthuysen1996}, its detection efficiency remained below 25\%, even if a non-balanced interferometer was employed. This efficiency barrier was broken by Kwiat \etal\ \cite{kwiat95} by using the quantum Zeno effect \cite{zeno77,Kofman2001}, i.e., the repeated weak interrogation \cite{Kwiat1998}. The same technique was further optimised and experimental efficiencies as high as 73\%  \cite{kwiat99} and 75\% \cite{Tsegaye1998} were obtained. A recent work based on quantum Zeno stabilisation of ultra-cold atoms \cite{Peise2015} has showed efficiencies of 90\%. The leap from an efficient IFM object detector to an imaging technique was taken in reference \cite{white98} where an actual implementation was reported.

IFM imaging can have a major impact in live-cell imaging \cite{Taylor2014, Cole2014} and low-damage biological imaging \cite{Taylor2013}. There are already detailed proposals to introduce this technique to electron microscopy \cite{Putnam2009, Kruit2016, Agarwal2019, Madan2020, Juffmann2017}. Here, not only extremely low probe beam fluxes are required \cite{Mitchison2001}, but one major problem is the spatial extension of microscopic objects \cite{Thomas2015}. 
These issues motivate our work.
 
We generalise the high-efficiency IFM imaging to spatially-extended objects by  simultaneously probing all the pixels without displacing and/or continuously re-aligning the sample. We propose a method using a superposition of photonic angular momentum (OAM) \cite{Allen1992,Leach2002} states, based on a mapping between each pixel and each OAM value.

The use of photons carrying OAM \cite{Barnett2017} stemmed from the Allen \etal\ proposal \cite{Allen1992} and nowadays is a mature technology with countless applications \cite{Tamburini2006, Boyd2011, Bozinovic2013, Xie2016}. Photonic OAM is due to the helical phase front along the propagation direction with quantized angular momentum $l\hbar$ with $l\in\mathbb{Z}$ \cite{Molina-Terriza2001}. The generation/detection of OAM states of light is done via Dove prisms \cite{Gonzalez2006}, refractive elements \cite{Lavery2012} or the more versatile spatial light modulators (SLM) \cite{Mirhosseini2013}.

The IFM schemes discussed above are formulated in terms of perfectly absorbing objects, while realistic scenarios involve semitransparent objects \cite{Jang1999, Garcia-Escartin2006,Mitchison2001,Thomas2014,
Azuma2006}.  In the latter case, single-pixel IFM schemes were found to reach the loss-free limit at a slower rate than for perfectly opaque objects \citep{Jang1999,Garcia-Escartin2006}. 
 It is then relevant to establish the  behaviour of the analogous IFM scheme for multi-pixel semitransparent objects.

The article is structured as follows. In Section \ref{sec:background} we briefly describe IFM experiments, from the Elitzur-Vaidman bomb tester to the  high-efficiency IFM employing the quantum Zeno effect. We introduce and discuss our Mach-Zehnder-based experimental proposal in Section \ref{sec:proposal}. In Section \ref{sec:mich} we optimize our proposal through the use of a Michelson interferometer and discuss the more realistic semi-transparent objects in Section \ref{sec:semitr}. We conclude in Section \ref{sec:conclusions}.

\section{From IFM experiments to efficient IFM imaging}
\label{sec:background}

\begin{figure}[]
\putfig{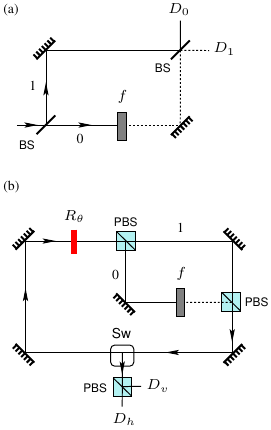}{6}
\caption{(a) Elitzur-Vaidman IFM  experimental setup; (b) high-efficiency IFM imaging scheme based on the quantum Zeno effect.}
\label{fig:EV_and_IFM_Zeno}
\end{figure}

The first IFM setup is due to Elitzur and Vaidman \cite{EV}. The experiment consists of a balanced Mach-Zehnder interferometer with an opaque object placed in one of the arms, see Fig.\ref{fig:EV_and_IFM_Zeno}(a). After the first beam-splitter, the photon is in an equal superposition of being in both arms, $0$ and $1$. If the object is absent, the photon interferes constructively at the second beam-splitter. Since the MZI is assumed balanced (50/50), $D_1$ never clicks while detector $D_0$ clicks with 100\% probability. However, the presence of an object $f$ destroys the interference and hence there is a non-zero probability for detector $D_1$ to click. A click in $D_0$ gives no information regarding the presence or absence of the object, however a click at detector $D_1$ signals the presence of the object, although the photon never interacted with it. For a single-photon source we have the seemingly paradoxical situation of detecting the object $f$ without ever interacting with it.

The photo-detection probabilities $p_i$ at detectors $D_i$ are shown in Table \ref{EV_simple}; $p_{abs}$ denotes the absorption probability and we use the convention $f=0\, (1)$ if the object is absent (present).

\begin{table}[ht]
\centering
\begin{tabular}{l|ccc}
\hspace{.1cm} $f$ & \hspace{.1cm} $p_0$ & \hspace{.1cm} $p_1$ & \hspace{.1cm} $p_{abs}$ \\ [0.5ex]
 \hline 
\hspace{.1cm} 0 & \hspace{.1cm} 1 & \hspace{.1cm} 0 & \hspace{.1cm} 0 \\ [0.5ex]
\hspace{.1cm} 1 & \hspace{.1cm} $\dfrac{1}{4}$ & \hspace{.1cm} $\dfrac{1}{4}$ & \hspace{.1cm} $\dfrac{1}{2}$
\end{tabular}
\caption{Probabilities of different outcomes in the Elitzur-Vaidman experiment.}
\label{EV_simple}
\end{table}

The Elitzur-Vaidman (EV) scheme has an efficiency of $25\%$; here we define the efficiency as the probability $p_1$ of the photon reaching $D_1$ if an object is present. To overcome the low efficiency of the EV experiment, Kwiat \etal\ proposed a scheme capable to achieve ideal efficiencies close to $100\%$ \citep{kwiat95, kwiat99}. The key to achieve this high efficiency is to use the quantum Zeno effect \citep{zeno77}: the coherent evolution of the photon is inhibited by a repeated weak measurement. In this case the quantum Zeno effect is based on an inhibited polarisation rotation.

The scheme is depicted in Fig.~\ref{fig:EV_and_IFM_Zeno}(b) and works as follows. A single photon is circulated $N$ times inside the cavity formed by the four mirrors, after which it is switched out (Sw) and its polarisation is measured. The photon is initially $H$-polarised and in each cycle $R_\theta$ rotates its polarisation with an angle $\theta= \frac{\pi}{2 N}$. An interferometer is defined by the two polarising beam-splitters (PBS). An opaque object is present (or absent) in the $V$-arm of the interferometer. If the object is absent, after $N$ cycles the polarisation is rotated from $H$ to $V$ and is thus detected with probability $p=1$ at the $D_v$ detector. However, if the object is present in the $V$-arm of the interferometer, the evolution from $H$ to $V$ is inhibited at each step. For each cycle, the probability that the photon is not absorbed is $p_{nabs} = \cos^2 \theta$ and in this case the photon is projected back to its initial polarisation state, $H$. After $N$ cycles, the probability to find the photon $H$-polarised is:
\begin{eqnarray} 
p_h= \prod_{n = 0}^{N - 1} p_{nabs} 
= \cos^{2N} \theta \approx  1- \frac{\pi^2}{4 N}
\label{pnabs0}
\end{eqnarray}
The absorbtion probability  $p_{abs}=1-p_h=\frac{\pi^2}{4 N}$ can be made arbitrarily small by increasing $N$. In the large $N$ limit, the probability to detect the photon in $D_h$ approaches 1, see Table \ref{HEFF}.

\begin{table}[ht]
\centering
\begin{tabular}{l|ccc}
\hspace{.1cm} $f$ & \hspace{.1cm} $p_h$ & \hspace{.1cm} $p_v$ & \hspace{.1cm} $p_{abs}$ \\ [0.5ex]
\hline 
\hspace{.1cm} 0 & \hspace{.1cm} 0 & \hspace{.1cm} 1 & \hspace{.1cm} 0 \\ [0.5ex]
\hspace{.1cm} 1 & \hspace{.1cm} $1-\dfrac{\pi^2}{4N}$ & \hspace{.1cm} $0$ & \hspace{.1cm} $\dfrac{\pi^2}{4N}$ \\
\end{tabular}
\caption{Detection probabilities for $D_h$ ($p_h$) and $D_v$ ($p_v$), and the absorption probability $p_{abs}$. We assume $N \gg 1 $.}
\label{HEFF}
\end{table}

\section{IFM imaging of multi-pixel objects}
\label{sec:proposal}

The previous schemes have an intrinsic limitation, namely they are able to detect/image a single object, or pixel. However, in real-life applications the sample of interest is an extended, multi-pixel object. In this section we extend the previous schemes to objects having multiple (transparent or absorbing) pixels. We first generalize the IFM scheme from Fig.\ref{fig:EV_and_IFM_Zeno}(a) to a multi-pixel object. We then use this approach to improve the efficiency by using the quantum Zeno effect, similar to the Kwiat \etal\ setup \cite{kwiat99}.

For the multi-pixel IFM imaging we use an extra degree of freedom (DoF) in which to encode the spatial structure of the sample. This DoF should be $d$-dimensional since we want to probe all the pixels simultaneously and we should be able to perform multiplexing/demultiplexing. Possible choices are path, wavelenght and OAM. The path DoF is not suitable in our scheme, since the information about each pixel will be lost after the multiplexing step and we will not gain any information about the object structure. Wavelength is not a good choice either, since the pixels can be opaque or transparent depending on the probing wavelenght (unless we convert all wavelengths to a given $\lambda_0$ before the object and convert them back after the object). These problems are avoided if we use OAM. The values OAM can take are theoretically unbounded and several sorting schemes for OAM are known \citep{OSullivan2012, Fu2018, sorter}. 

\begin{figure}[]
\putfig{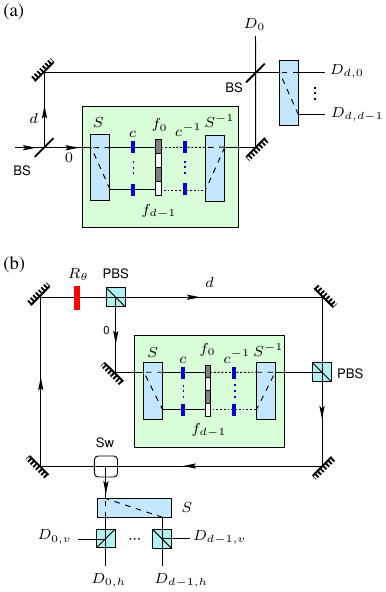}{8}
\caption{IFM imaging of a multi-pixel object. (a) single-pass interferometer; (b) high-efficiency, multi-pass interferometer scheme.}
\label{general}
\end{figure}

\noindent {\em (i) Generalised IFM imaging.} In order to encode the spatial information of a multi-pixel object in the internal DoF (OAM), we need to:\\
(a) initialise the single photon in an equal superposition of all OAM states. This is necessary since we want to explore all the pixels in parallel with a single photon. Clearly, the dimension of the OAM space has to be the same as the number of pixels;\\
(b) transfer the information between spatial and OAM Dof.

Our scheme is shown in Fig.~\ref{general}(a). Similar to Fig~\ref{fig:EV_and_IFM_Zeno}(a), the object is situated in the 0-arm of a balanced Mach-Zehnder interferometer, but in this case it is inside a path-to-OAM encoder (green box). The encoder transfers the information from the $\ell$-th pixel (transparent/opaque, i.e., $0/1$) to the corresponding OAM value $\ket{\ell}_{OAM}$.

We label the two arms of the MZI as $0$ and $d$; the spatial modes $0$ to $d-1$ correspond to the sample's pixels. The photon enters the interferometer in spatial mode $0$ and the first beam-splitter places it in a superposition of the two spatial modes $0$ and $d$. Before reaching the object in arm $0$, a sorter $S$ demultiplexes the OAM values $\ell$ to different paths corresponding to the pixels. We take $f_\ell= 0\, (1)$ if the $\ell$-th pixel is transparent (opaque). After the interaction with the object, all OAM components are multiplexed back to the same path by an inverse sorter $S^{-1}$.

Finally, all OAM modes interfere at the second beam-splitter. As before, a click in the detector $D_0$ gives us no information about the object. The information about the $\ell$-th pixel is encoded in the OAM mode $\ell$ which exists on the port $d$. Therefore on path $d$ we demultiplex again the OAM values $\ell$ to different detectors $D_{d,\ell}$. A click in the detector $D_{d, \ell}$ tells us that the $\ell$-th pixel was opaque, since only in this case the constructive interference at the second beam-splitter was destroyed.

It is important to note that a photon with OAM mode $\ell$ has a radial extension proportional to $\ell$. Thus, in order to probe all pixels with the same transversal Gaussian mode, i.e., $\ket{0}_{OAM}$, we insert OAM mode converters $c$ ($c^{-1}$, its inverse) before (after) the object. On each path $\ell$ the OAM converter $c$ changes $\ket{\ell}_{OAM} \mapsto \ket{0}_{OAM}$, and $c^{-1}$ performs the inverse transformation. These are spiral phase-plates of order $\ell$, together with their inverses.

\begin{figure}[]
\putfig{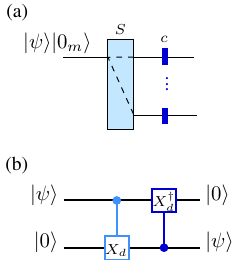}{3.5}
\caption{(a) Implementation of a SWAP gate between OAM and path DoFs for states $\ket{\psi}\ket{0}$, i.e., when all photons enter in the 0-th path. (b) Equivalent quantum network for the SWAP gate when the path qudit is in the $\ket{0}$ state: a $C(X_d) (\mathrm{OAM, path})$ gate followed by a $C(X_d^\dag) (\mathrm{path, OAM})$ gate \citep{mermin2001, garcia2013}.}
\label{swap}
\end{figure}

The combination between the sorter $S$ and converter $c$ is equivalent to a qudit SWAP gate between OAM and path degrees of freedom, Fig.~\ref{swap}. Thus, after interacting with the object, the spatial information about the sample (on/off pixels) contained in the path qudit is swapped into the OAM qudit by $c^{-1}$ and $S^{-1}$.

The probabilities for a detector to click are shown in Table \ref{EV_general}, see Appendix \ref{nabs EV}. Notice that they are equal to the probabilities of the EV experiment (Table \ref{EV_simple}) scaled by $1/d$, where $d$ is the dimension of OAM (or the total number of pixels). One can understand this by observing that the setup in Fig.~\ref{general}(a) is equivalent to $d$ Elitzur-Vaidman experiments run in parallel.

{\renewcommand{\arraystretch}{2}
\begin{table}[ht]
\centering
\begin{tabular}{l|ccc}
\hspace{.1cm} $f_{\ell}$ & \hspace{.1cm} $p_{0,\ell}$ & \hspace{.1cm} $p_{d,\ell}$ & \hspace{.1cm} $p_{abs}$ \\ [0.5ex]
\hline 
\hspace{.1cm} 0 & \hspace{.1cm} $\dfrac{1}{d}$ & \hspace{.1cm} 0 & \hspace{.1cm} 0 \\ [1.5ex]
\hspace{.1cm} 1 & \hspace{.1cm} $\dfrac{1}{4 d}$ & \hspace{.1cm} $\dfrac{1}{4 d}$ & \hspace{.1cm} $\dfrac{1}{2 d}$
\end{tabular}
\caption{Probabilities of different outcomes for the multi-pixel IFM experiment.}
\label{EV_general}
\end{table}
}

\noindent {\em (ii) Generalized high-efficiency imaging.} As before, we can improve the efficiency of the interaction-free measurement by using the quantum Zeno effect. The multi-pixel generalization of the Kwiat \etal\ scheme \cite{kwiat95, kwiat99} is shown in Fig.~\ref{general}(b). The main idea is similar: a photon with $H$-polarisation performs multiple cycles of the setup. At the beginning of each cycle the polarisation is rotated by a small angle $\theta = \frac{\pi}{2 N}$, the photon being now in a superposition of $H$ and $V$ polarisations. The two polarisation components are sorted on spatial modes $d$ and $0$ by a polarising beam-splitter. The object is situated in the $V$-arm.

As before, we use the photonic OAM to encode the information about the pixels of the sample by inserting a sorter $S$ and an inverse sorter $S^{-1}$ before and after the object respectively; the converters $c, c^{-1}$ ensure that all pixels are probed with a Gaussian, $\ell=0$ OAM mode. If a pixel $\ell$ is transparent, the polarisation is rotated stepwise towards $V$ and the detector $D_{\ell,v}$ will click after $N$ cycles. If the pixel $\ell$ is opaque, the probability for the photon to be absorbed is small because the $V$-polarisation component is small. At each cycle the photon state is projected, with a large probability, on the $H$-polarisation. The probability of transmission after $N$ cycles for a photon with $H$-polarisation is (see Appendix \ref{nabs kwiat}):
\begin{equation} 
\prod_{n = 0}^{N - 1} p_{nabs} 
\approx 1- \frac{N_{abs}}{d}\frac{\pi^2}{4 N}
\label{pnabs1}
\end{equation}
where $N_{abs} = \sum_{\ell=0}^{d-1} f_\ell $ is the number of opaque pixels and $d$ is the total number of pixels. In this case the detector $D_{\ell,h}$ will click. For each OAM value $\ell$ the presence of an opaque (transparent) pixel leads to a click in the corresponding detector $D_{\ell,h}$ ($D_{\ell,v}$) due to the quantum Zeno effect.

{\renewcommand{\arraystretch}{2}
\begin{table}[ht]
\centering
\begin{tabular}{l|ccc}
\hspace{.1cm} $f_\ell$ & \hspace{.1cm} $p_{\ell,h}$ & \hspace{.1cm} $p_{\ell,v}$ & \hspace{.1cm} $p_{abs}$\\ [0.5ex]
\hline 
\hspace{.1cm} 0 & \hspace{.1cm} 0 & \hspace{.1cm} $\dfrac{1}{d}$ & \hspace{.1cm} 0 \\ [0.5ex]
\hspace{.1cm} 1 & \hspace{.1cm} $\dfrac{1}{d}\left(1-\dfrac{\pi^2}{4 N}\right)$ & \hspace{.1cm} 0 & \hspace{.1cm} $\dfrac{1}{d}\dfrac{\pi^2}{4 N}$
\end{tabular}
\caption{Probabilities for the photon to reach one of the detectors or to be absorbed after $N$ cycles for the case of an opaque multi-pixel object.}
\label{HEFF_generalized}
\end{table}
}

\section{Michelson setup}
\label{sec:mich}

\begin{figure}[]
\putfig{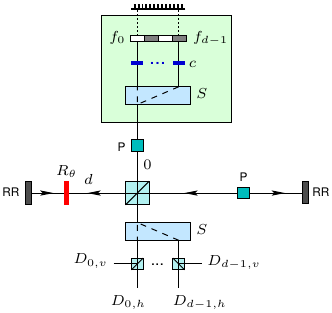}{7}
\caption{Michelson configuration for interaction-free imaging of a multi-pixel object: high-efficiency/multi-pass scheme.}
\label{michel}
\end{figure}

We can simplify the high-efficiency scheme if we use a Michelson interferometer. This setup requires only two sorters, compared to three in the Mach-Zehnder case.

In the folded high-efficiency configuration Fig.~\ref{michel} the photon undergoes $N$ cycles, similar to the scheme in Fig.~\ref{general}(b). The photon enters on path $d$ and successively passes through a polarisation rotator. It is reflected back from the retro-reflector $RR$ and after the second pass through the polarisation rotator a polarising beam-splitter directs the photon on spatial modes $0$ or $d$ according to its polarisation. The subsequent evolution of the photon state is the same as in the Michelson/Elitzur-Vaidman experiment until the photon reaches again the polariser beam-splitter and starts a new cycle. Since the photon passes through the polarisation rotator twice during a cycle, the rotation angle is now $\theta = \frac{\pi}{4 N}$.

On path $d$ we use retro-reflectors, such that the photon has a double reflection and the OAM state $\ket{\ell}$ is unchanged; a mirror reflection is equivalent to $\ket{\ell} \mapsto \ket{-\ell}$. On path $0$ we have a single mirror, since the photon is in $\ell=0$ state after the converter $c$.

To switch out the photon after $N$ cycles, we use Pockels cells $P$ placed in both arms. When activated, the Pockels cells rotate the polarisation by 90$\degree$, and thus photons from both arms exit through the same port of the PBS. In this case the information is reversed: clicks in $D_{\ell,v}$ ($D_{\ell,h}$) tell us the pixel $\ell$ is opaque (transparent).

\section{Semi-transparent objects}
\label{sec:semitr}

So far we have discussed objects with fully opaque or fully transparent pixels. We now extend our proposal to semi-transparent objects, which are better suited to describe realistic cases like biological tissues, an important application of IFM imaging.

We consider a muti-pixel object and denote by $T_\ell$ the transmission coefficient of pixel $\ell$. The action of the path-to-OAM encoder is described by an OAM-dependent transmission coefficient matrix, see Appendix \ref{semit}. For a single cycle the transformation on the photon is $\mathcal{M} = \mathcal{OR}$, where $\mathcal{O}$ and $\mathcal{R}$ are the action of the interferometer and the rotator respectively. The total transformation after $N$ cycles is
\begin{equation}
\mathcal{M}^N = \diag (m_0^N,m_1^N,\dots,m_{d-1}^N)
\end{equation}
where each block 
\begin{equation}
m_\ell = 
	\begin{bmatrix}
	\cos{\theta} & -\sin{\theta} \\
	\sqrt{T_\ell}\sin{\theta} & \sqrt{T_\ell}\cos{\theta} 
	\end{bmatrix}
\end{equation}
encodes the combined effects of polarisation rotation and transmission through the semitransparent object for OAM value $\ell$. By applying the transformation $\mathcal{M}^N$ on the initial state, we obtain the probabilities for the photon to reach the detectors $D_{\ell,h}, D_{\ell,v}$
\begin{eqnarray}
\label{hsemi}
p_{\ell,h} &=& \dfrac{1}{d}\left(1 - \dfrac{1+\sqrt{T_\ell}}{1-\sqrt{T_\ell}} \dfrac{\pi^2}{4N} \right) \\
\label{vsemi}
p_{\ell,v} &=& \dfrac{1}{d} \dfrac{T_\ell}{(1-\sqrt{T_\ell})^2} \dfrac{\pi^2}{4N^2} 
\end{eqnarray}

Remarkably, the absorbtion probability $p_{abs}= 1- p_{\ell,h}- p_{\ell,v}$ vanishes for large $N$, so in this case we can still talk about interaction-free measurements. As expected for semi-transparent pixels, the probability for the photon to be detected in any $D_{\ell,v}$ detector is nonzero. However, this probability scales as $1/N^2$ and thus the photon is more likely to be detected in a $D_{\ell,h}$ detector. Also, for semi-transparent objects reaching the lossless limit is slower for large $N$.

\section{Discussion}
\label{sec:conclusions}

Current imaging techniques have several limitations in terms of resolution, spectral range and/or contrast. One way to overcome these limitations is to use various quantum-enhanced techniques, like sub-shot-noise imaging \cite{Treps2002, Brida2010, Samantaray2017} or imaging with undetected photons \cite{Lemos2014, Kalashnikov2016}. A target application for our IFM setup are extremely sensitive samples requiring imaging in low-light conditions. In this article we have extended IFM imaging to semi-transparent, multi-pixel samples. This generalization keeps all the advantages of previous schemes \citep{EV, kwiat95, kwiat99}.

First, we have modeled the sample as a collection of fully transparent/opaque pixels. In our scheme we encode the information about the sample into OAM of single photons. By starting with a photon in a superposition of OAM states, we can probe multiple pixels in parallel.

A central ingredient of our proposal is the quantum sorter, which demultiplexes the photon according to its orbital angular momentum and allows us to encode the information about the pixels. The quantum sorter is also essential to recover the information at the end by directing the photon to the corresponding detector. Clearly, the number of detectors should be the same as the number of pixels. State-of-the-art single-photon countig cameras can have up to 512$\times$512 pixels, where each pixel is an independent single-photon detector \citep{Ulku2017, Bruschini2019}.

We have discussed two setups. In a Mach-Zehnder configuration we need three OAM $d$-mode sorters and two converters, whereas for a Michelson setup this reduces to only two OAM sorters and one converter. This is a consequence of the fact that both the sorter $S$ and the converter $c$ act as their own inverses if the photon enters from the opposite direction. Moreover, if the acquisition speed is not very important, the final OAM sorter can be replaced by a computer-controlled SLM \cite{Mirhosseini2013} and thus only two photo-detectors are needed, one for each polarisation.

In the case of IFM imaging of perfectly opaque (or transparent) pixels, the absorption probability is lower compared to the single-pixel case. This is due to the photon evolving coherently in a superposition of paths, leading to the factor $N_{abs}/d$ in Eq.~(\ref{pnabs1}). In the large $N$ limit, we obtain the same probabilities as in the single-pixel case.

In this paper we have discussed only the case of ideal, lossless components. However, for real systems the achievable efficiencies have stronger constraints. The intrinsic losses reduce the real efficiency compared to the ideal one, as the number of cycles $N$ increases, the former being expected to reach a maximum value $<1$ at a finite number of cycles \citep{kwiat99}. 

Interaction-free imaging schemes can also be implemented with weak coherent states, which are experimentally more accesible than Fock states \citep{kwiat99}. In this case we need to reduce optical imperfections and interferometric instabilities to reach an efficiency close to 1.

For real-life applications we need to consider semi-transparent pixels, where each pixel $\ell$ has a transmission coefficient $T_\ell\in [0,1)$. Similar to the single-pixel case \citep{Jang1999, Azuma2006}, we find that for multiple semi-transparent pixels we can still reach the lossless limit, but at a slower rate compared to a fully opaque object, eqs.~\eqref{hsemi}, \eqref{vsemi}. While interaction-free measurements do not improve over classical ones when trying to determine an unknown transparency, they do allow to distinguish between high-contrast semitransparent samples \citep{Thomas2014}. In our parallel setup, this can be realised at the single-pixel level by analysing the clicks in the corresponding detectors.

Although here we have discussed 1D samples, conceptually our method can be extended to 2D objects as well. In order to image a 2D sample we can use another photonic degree of freedom, e.g., the radial quantum number $r$. By demultiplexing the photon in the second dimension according to the second DoF, we can describe each pixel by a pair $(\ell, r)$. In this sense, there are promising sorting schemes for the radial quantum number \citep{zhou2017, gu2018}, together with equivalent spiral-phase plates \citep{Ruffato2014, Ruffato2015}.

\begin{acknowledgments}
The authors acknowledge support from a grant of the Romanian Ministry of Research and Innovation, PCCDI-UEFISCDI, project number PN-III-P1-1.2-PCCDI-2017-0338/79PCCDI/2018, within PNCDI III. R.I.~acknowledges support from PN 19060101/2019-2022.
S.A. also acknowledges support by the Extreme Light Infrastructure 
Nuclear Physics (ELI-NP) Phase II, a project co-financed by the Romanian 
Government and the European Union through the European Regional 
Development Fund and the Competitiveness Operational Programme 
(1/07.07.2016, COP, ID 1334).
\end{acknowledgments}

\begin{widetext}
\appendix
\section{Interaction-free imaging, multi-pixel case}
\label{nabs EV}

Here we calculate the probabilities in Table \ref{EV_general}, see Fig.~\ref{general}(a). For simplicity we omit the action of the OAM converters ($c, c^{-1}$) situated immediately before and after the object, since they do not change the calculations. The total Hilbert space is $\mathcal{H}_{tot} = \mathcal{H}_{O} \otimes \mathcal{H} _{m}$, where $\mathcal{H}_{O}$ and $\mathcal{H} _{m}$ are the Hilbert spaces corresponding to the OAM and spatial mode, respectively. The photon is initially in the state 
\begin{equation}
\vert \psi_{0} \rangle  = \frac{1}{\sqrt{d}}\sum_{\ell=0}^{d-1}\vert \ell \rangle \otimes \vert 0_m \rangle 
\end{equation}
After the first beam-splitter (equivalent to a Hadamard gate on the spatial mode) the state becomes
\begin{equation}
\vert \psi_{1} \rangle = \frac{1}{\sqrt{2d}} 
\left(  \sum_{\ell=0}^{d-1} \vert \ell \rangle \vert 0_m \rangle 
+ \sum_{\ell=0}^{d-1} \vert \ell \rangle  \vert d_m \rangle \right)
\end{equation}
The sorter $S$ demultiplexes the photons into different spatial modes according to their OAM:
\begin{equation}
\vert \psi_{2} \rangle = \frac{1}{\sqrt{2d}} \left(  \sum_{\ell=0}^{d-1} \vert \ell \rangle \vert \ell_m \rangle + \sum_{\ell=0}^{d-1} \vert \ell \rangle  \vert d_m \rangle \right)
\end{equation}
The total probability of the photon to be absorbed by any of the pixels is given by
\begin{eqnarray} \nonumber
p_{abs} &=& \langle \psi_2 \vert \left( \sum_{\ell=0}^{d-1} f_\ell \, \vert \ell,\ell_m \rangle \langle \ell_m, \ell \vert \right)\vert \psi_2 \rangle 
\\
&=& \frac{N_{abs}}{2d} 
\end{eqnarray}
where $N_{abs} = \sum_{\ell=0}^{d-1} f_\ell$ is the number of opaque pixels. The probability for the photon to be transmitted is then
\begin{equation}
\label{pneabs}
p_{nabs} = 1 - p_{abs} = \frac{2d - N_{abs}}{2d}  
\end{equation}
In the case of non-absorption, after the inverse sorter $S^{-1}$ the state is
\begin{equation}
\vert \psi_{3} \rangle = \frac{1}{\sqrt{2d-N_{abs}}} \left[ \sum_{\ell=0}^{d-1} (1-f_\ell) \, \vert \ell \rangle \vert 0_m \rangle +  \sum_{\ell=0}^{d-1} \vert \ell \rangle \vert d_m \rangle \right]
\end{equation}
and after the second beam splitter the output state becomes
\begin{eqnarray} \nonumber
\label{psif}
\vert \psi_{f} \rangle &=& \frac{1}{\sqrt{2d-N_{abs}}} \left[ \sum_{\ell=0}^{d-1} (1-f_\ell) \, \vert \ell \rangle \frac{1}{\sqrt{2}} (\vert 0_m \rangle + \vert d_m \rangle) + \sum_{\ell=0}^{d-1} \vert \ell \rangle  \frac{1}{\sqrt{2}} (\vert 0_m \rangle - \vert d_m \rangle) \right]\\
&=& \frac{1}{\sqrt{2d-N_{abs}}} \left\{ \frac{1}{\sqrt{2}}  \left[\sum_{\ell=0}^{d-1} (1-f_\ell) \, \vert \ell \rangle + \sum_{\ell=0}^{d-1} \vert \ell \rangle \right] \vert 0_m \rangle + \frac{1}{\sqrt{2}}  \left[\sum_{\ell=0}^{d-1} (1-f_\ell) \, \vert \ell \rangle - \sum_{\ell=0}^{d-1} \vert \ell \rangle \right] \vert d_m \rangle \right\}
\end{eqnarray}
From eq.~\eqref{pneabs} and \eqref{psif} we obtain the probabilities in Table \ref{EV_general}.

\section{Interaction-free imaging of a multi-pixel object, high-efficiency experiment}
\label{nabs kwiat}

In this case we have an extra degree of freedom, the polarisation. The total Hilbert space is $\mathcal{H}_{tot} = \mathcal{H}_{p} \otimes \mathcal{H} _{O} \otimes \mathcal{H} _{m}$, where $\mathcal{H}_{p}$, $\mathcal{H}_{O}$ and $\mathcal{H} _{m}$ are the Hilbert spaces corresponding to the polarisation, OAM and spatial mode, respectively. The photon starts in the initial state:
\begin{equation}
\label{psiinOAM}
\vert \psi_{in} \rangle  = \vert H \rangle \otimes \frac{1}{\sqrt{d}}\sum_{\ell=0}^{d-1}\vert \ell \rangle \otimes \vert d_m \rangle
\end{equation}

The photon undergoes a cyclic evolution inside the loop until it is switched out after $N$ cycles. For each cycle, we have a polarisation rotation $R_\theta$, followed by a PBS, sorters $S$, $S^{-1}$ and the second PBS. After $n$ cycles the state of the photon is:
\begin{equation}
\label{psigen}
\vert \psi' \rangle  = \frac{1}{\mathcal{N}_f} \left\{ \cos^n \theta \vert H \rangle \sum_{\ell=0}^{d-1} f_\ell \,\vert \ell \rangle  + [\sin(n \theta) \vert V \rangle + \cos(n \theta) \vert H \rangle ]  \sum_{\ell=0}^{d-1} (1-f_\ell) \, \vert \ell \rangle   \right\} \vert d_m \rangle
\end{equation}

Let $N_{abs} = \sum_{\ell=0}^{d-1} f_\ell $ be the number of opaque pixels. Then from eq.~\eqref{psigen} we have:
\begin{equation}
\mathcal{N}_f = \sqrt{d-N_{abs}+N_{abs}\cos^{2n}\theta }
\end{equation}

In order to calculate the absorption probability during an arbitrary cycle, we consider another iteration starting from the state \eqref{psigen}. Immediately after the polarisation rotator the state is
\begin{eqnarray} \nonumber
\vert \psi'' \rangle  &=& \frac{1}{\mathcal{N}_f} \left\{( \cos^{n+1} \theta \vert H \rangle + \cos^n \theta \sin \theta \vert V \rangle ) \sum_{\ell=0}^{d-1} f_\ell \, \vert \ell \rangle  \right.\\
&+& \left.[\sin((n+1) \theta) \vert V \rangle + \cos((n+1) \theta) \vert H \rangle ] \sum_{\ell=0}^{d-1} (1-f_\ell) \, \vert \ell \rangle   \right\}  \vert d_m \rangle
\end{eqnarray}

The absorption probability for the $(n+1)$-th cycle is:
\begin{equation}
p_{abs}(n) = \frac{N_{abs}}{\mathcal{N}_f^{\,2}} \cos^{2n}\theta \sin^2 \theta =\frac{N_{abs} \cos^{2n}\theta \sin^2 \theta }{d-N_{abs}+ N_{abs} \cos^{2n}\theta}
\end{equation}

The non-absorption probability for the $(n+1)$-th cycle is then:
\begin{equation}
\label{Nf}
p_{nabs}(n) = \frac{d-N_{abs}+ N_{abs} \cos^{2n+2}\theta}{d-N_{abs}+ N_{abs} \cos^{2n}\theta} 
\end{equation}

The non-absorption probability after $N$ completed cycles:
\begin{eqnarray} \nonumber
p_{nabs}= \prod_{n = 0}^{N - 1} p_{nabs}(n) 
&=& 1 - \frac{N_{abs}}{d} (1-\cos^{2N}\theta) \\
& \approx & 1 - \frac{N_{abs}}{d} \frac{\pi^2}{4N}
\end{eqnarray}

For $N \gg 1$, the non-absorption probability $p_{nabs} \rightarrow 1$. For $\theta = \frac{\pi}{2N}$ the final state in eq. (\ref{psigen}) is:
\begin{equation}
\label{psifinal}
\vert \psi_{f} \rangle = \frac{1}{\sqrt{d}} \left[\vert H \rangle  \sum_{\ell=0}^{d-1} f_\ell \, \vert \ell \rangle  + \vert V \rangle \sum_{\ell=0}^{d-1} (1-f_\ell) \, \vert \ell \rangle   \right]  \vert d_m \rangle
\end{equation}
as in Table \ref{HEFF_generalized}.

\section{Interaction-free imaging of a semi-transparent multi-pixel object}
\label{semit}

In this section we discuss the general case of semi-transparent multi-pixel objects. This generalises Jang's approach \citep{Jang1999} used for polarisation degree of freedom. In our case the total Hilbert space is $\mathcal{H}_{tot} = \mathcal{H} _{O} \otimes \mathcal{H}_{p}$ where $\mathcal{H}_{O}$ and $\mathcal{H}_{p}$ are the Hilbert spaces corresponding to the OAM and polarisation respectively. For simplicity we omit the spatial mode Hilbert space and model the action of the path-to-OAM encoder with an OAM dependent transmission coefficient matrix
\begin{equation}
\mathcal{T}_{OAM} = \diag (\sqrt{T_0},\sqrt{T_1},\dots,\sqrt{T_{d-1}})
\end{equation}
This is used to define the total action of the interferometer by
\begin{equation}
\mathcal{O} = \mathbb{I}_d \otimes \begin{bmatrix}
1 & 0 \\
0 & 0 
\end{bmatrix}
 +  \mathcal{T}_{OAM} \otimes  \begin{bmatrix}
0 & 0 \\
0 & 1 
\end{bmatrix}
\end{equation}
where $\mathbb{I}_d$ is the identity on the OAM space. The polarisation rotator is
\begin{equation}
\mathcal{R} = \mathbb{I}_d \otimes \begin{bmatrix}
\cos{\theta} & -\sin{\theta} \\
\sin{\theta} & \cos{\theta} 
\end{bmatrix}
\end{equation}
Then the total transformation has a block-diagonal form
\begin{equation}
\mathcal{M} = \mathcal{OR}= \diag (m_0, m_1,\dots, m_{d-1})
\end{equation}
where each block 
\begin{equation}
m_\ell = 
	\begin{bmatrix}
	\cos{\theta} & -\sin{\theta} \\
	\sqrt{T_\ell}\sin{\theta} & \sqrt{T_\ell}\cos{\theta} 
	\end{bmatrix}
\end{equation}
represent the combined effects of polarisation rotation and transmission through the semitransparent object for the OAM value $\ell$. The total transformation after $N$ cycles is also block-diagonal 
\begin{equation}
\mathcal{M}^N = \diag (m_0^N,m_1^N,\dots,m_{d-1}^N)
\end{equation}
which generalizes the results of Jang \citep{Jang1999} and Azuma \citep{Azuma2006} for the single-pixel case.

We apply the total transformation to the initial state 
\begin{equation}
\vert \psi_{in} \rangle  = \frac{1}{\sqrt{d}} [\,1 ~~1~\dots~ 1\,]^T \otimes [\,1 ~~ 0\,]^T= \frac{1}{\sqrt{d}} [\,1 ~~0~~1~~0~\dots~1~~ 0\,]^T
\end{equation}
and obtain

\begin{equation}
\vert \psi_{f} \rangle  =  \mathcal{M}^N \vert \psi_{in} \rangle  = [\,c_{0,h} ~~c_{0,v}~~c_{1,h} ~~c_{1,v}~\dots~c_{d-1,h} ~~c_{d-1,v}\,]^T
\end{equation}
where $c_{\ell,h}$ and $c_{\ell,v}$ correspond to the single-pixel amplitudes given in eq.~(13) of \citep{Jang1999}. The total probability of nonabsorbtion for each polarisation is
\begin{equation} 
\label{phsemi}
p_h=\sum_{\ell=0}^{d-1} |c_{\ell,h}|^2
\approx  1 - \frac{1}{d} \sum_{\ell=0}^{d-1}\frac{1+\sqrt{T_\ell}}{1-\sqrt{T_\ell}} \frac{\pi^2}{4N},
\end{equation}

\begin{equation} 
\label{pvsemi}
p_v =\sum_{\ell=0}^{d-1} |c_{\ell,v}|^2
\approx \frac{1}{d} \sum_{\ell=0}^{d-1} \frac{T_\ell}{(1-\sqrt{T_\ell})^2} \frac{\pi^2}{4N^2}
\end{equation}
for $N\gg 1$, $\theta = \frac{\pi}{2N}$ and $T_\ell \in [0,1)$. For the single-pixel case we recover the results of Ref.~\citep{Azuma2006}. 

\end{widetext}

%
%
\bibliographystyle{apsrev4-1}

\bibliography{articol_imaging}

\end{document}